# On spacetime coordinates in special relativity

Nilton Penha[*]
Departamento de Física, Universidade Federal de Minas Gerais, Brazil.

Bernhard Rothenstein[**]
"Politehnica" University of Timisoara, Physics Department, Timisoara, Romania.

Starting with two light clocks to derive time dilation expression, as many textbooks do, and then adding a third one, we work on relativistic spacetime coordinates relations for some simple events as emission, reflection and return of light pulses. Besides time dilation, we get, in the following order, Doppler k-factor, addition of velocities, length contraction, Lorentz Transformations and spacetime interval invariance. We also use Minkowski spacetime diagram to show how to interpret some few events in terms of spacetime coordinates in three different inertial frames.

## I – Introduction

Any physical phenomenon happens independently of the reference frame we may be using to describe it. It happens at a spatial position at a certain time. Then to characterize one such event we need to know three spatial coordinates and the time coordinate in a four dimensional spacetime, as we usually refer to the set of all possible events. In such spacetime, an event is a point with coordinates *(x,y,z,t) or (x,y,z,ct)* as it is much used where c is the light speed in the vacuum, admitted as constant according to the second postulate of special relativity; the first postulate is that "all physics laws are the same in all inertial frames".

Treating the time as *ct* instead of *t* has the advantage of improving the transparency of the symmetry that exists between space and time in special relativity. We can see this clearly in the Lorentz transformations equations. It worth mention the famous assertive[1] from Herman Minkowski: "Henceforth space by itself , and time by itself, are doomed to fade away into mere shadows, and only a kind of union of the two will preserve an independent reality".

Obviously, one given event has different coordinates depending on the reference frame that is used. The event by itself does not determine which other events are simultaneous to it; as in ordinary school geometry a given point does not determine by itself which other points lie behind it or to its left. Such specifications - *behind* and *left* - only make sense if we choose directions to which we can relate *behind* and *left*. In spacetime, the set of events which are simultaneous to a given event is determined only if one specifies the observer for whom simultaneity is to hold.

## II - Einstein's clocks synchronization

Einstein, in his analysis of simultaneity [2] provides a prescription for synchronizing clocks by using light rays which propagate with constant speed no matter the direction they go. The operational procedure consists in placing all clocks – not running yet clocks - at fixed grid points and choosing one of them as a master clock, usually the one in the spatial origin of the reference inertial frame. The master clock should be set to start at an arbitrary $ct_0$ and all the others set to start reading $ct = ct_0 + x$, where x is distance from the master clock.

Initially all clocks are stopped. Then a light source at the master clock's grid location emits a pulse and starts running and showing $ct_0$. As the pulse travels along the grid points it triggers off each one of the clocks which starts reading $ct = ct_0 + x$ as settled. Assuming that clocks are all alike they all will be showing the same time after that; we say they are synchronized. In Figure 1 we try to show schematically how things happen during the synchronization process.

---

[*] nilton.penha@gmail.com
[**] brothenstein@gmail.com

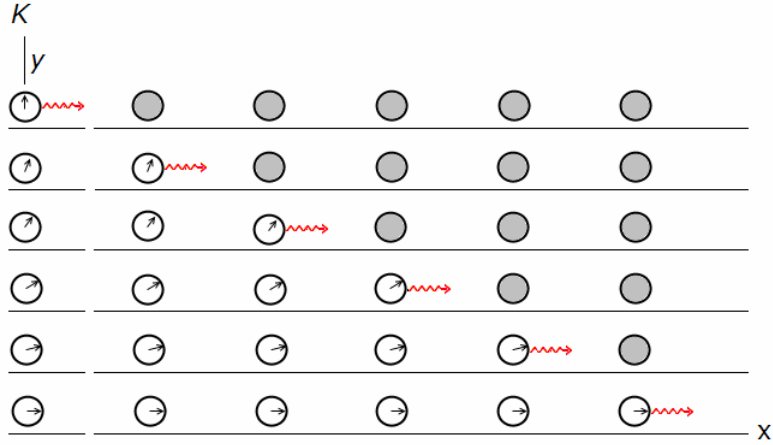

**Figure 1: The clocks are displayed as a grid along x-*axis* and each clock is prepared to start running at $ct = ct_0 + x$, when the light pulse, coming from the master clock at the origin and at $ct_0$ reaches its spatial position.**

## III - The light clock and time dilation

The light clock is a pedagogical device used by many authors for deriving the formula that accounts for the time dilation relativistic effect [3,4].

A light clock can be imagined as consisting of two mirrors $M_1$ and $M_2$ facing each other in an evacuated rigid box. Let the mirrors be at rest in a inertial frame $K=K(x,y,ct)$. As shown Figure 2a, $M_1$ and $M_2$ are placed along the spatial direction *y-axis*, being $M_1$ at the origin of $K$; the spatial distance between them is $d$, measured in $K$. A short light pulse propagates between the mirrors, and at $M_1$ there is a detector that counts the number of arriving reflections. If we manage to keep this device working, i.e., light going back and forth between $M_1$ and $M_2$, we have a light clock. The period of such clock is given by

$$c\tau = 2d .\qquad(1)$$

At this point we should recall the definition of *proper time*. Proper time, in special relativity, is the time measured by a single clock between events that occur at the same place as the clock. So $\tau$ is a proper time for the clock in $K$.

Let us have also a second light clock, identical to the first one, at rest in a second inertial frame $K'=K'(x',y',ct')$; the configuration of this light clock is just identical to first one: mirror $M_1'$ is at the origin of $K'$ and mirror $M_2'$ is at a distance $d' = d$, by construction, along the *y'-axis*. Let us consider the two reference frames in the so called standard configuration: *y-axis* and *y'-axis* coincide with each other and also *x-axis* and *x'-axis* are coincident and $K'$ is moving with constant dimensionless speed $\beta(V) \equiv V/c$ with respect to $K$ along the the *x-axis* toward positive values of $x$; also $ct = ct' = 0$ when the spatial origins of both inertial frames coincide with each other: $(x,y,ct) = (0,0,0) = (x',y',ct')$. The period of such light clock is $c\tau' = 2d'$, by construction. So in their own proper frame (each light clock is at rest in its own reference frame $K$ and $K'$ respectively), the light clocks have the same proper period ($c\tau' = c\tau$).

From the Figure 2b, by using Pythagoras' theorem, we easily show that

$$\left(\frac{cT}{2}\right)^2 = d^2 + \left(\frac{VT}{2}\right)^2, \qquad d = \frac{c\tau'}{2}, \tag{2}$$

$$cT = \gamma(V)c\tau' \tag{3}$$

where

$$\gamma(V) = \frac{1}{\sqrt{1-\beta(V)^2}} \tag{4}$$

and $cT$ is the period of the moving light clock at the viewpoint of the rest frame $K$. Since $\gamma(V) > 1$, $cT$ is bigger than $c\tau' = c\tau$. This means that the period that the observers in $K$ ascribe to the moving clock is longer; in other words, for the observer in $K$, time goes slower in the moving inertial frame $K'$. The time $cT$ is certainly not a proper time since it refers to two events that occur at different spatial points. In this case $cT$ is simply called a *time coordinate*. One frequently, among the physicist, refers to the proper time as being the time measured by observer's wristwatch, assuming no movement of it with respect to the observer.

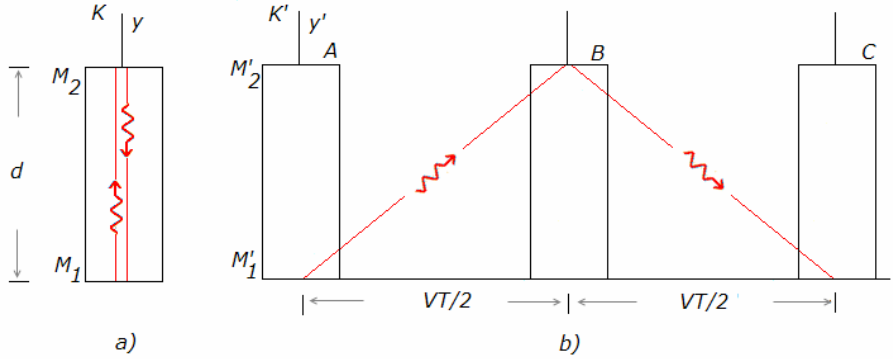

**Figure 2:  a) The light pulse going up and down assures a period $c\tau = 2d$;**
**b) The light clock moves to the right with dimensionless speed $\beta(V)$; observers at rest in $K$ ascribe to the moving clock a period $cT > c\tau' = c\tau$.**

We may admit that at each clock spatial position there is an imaginary observer, who reads the times shown by their own clocks (wristwatches). Observers in $K$ read the time at the grid point positions whose clocks have already been synchronized; observer in $K'$ reads the time at its origin space position clock.

By following the movement of the light clock in $K'$, in which it is at rest, we verify that the spatial separation between $K$ and $K'$ origins is $x = Vt$, when it shows a proper time $ct'$ and the clock in $K$ which is side by side with the light clock in $K'$ shows time $ct$. These two events are considered simultaneous by observers in K. According to the discussion above $ct$ and $ct'$ are related by

$$ct = \gamma(V)ct'. \tag{5}$$

Since both frames have the same spacetime origin, we can, with no loss of generality, write that

$$c\Delta t = \gamma(V)c\Delta t', \tag{6}$$

where $c\Delta t$ and $c\Delta t'$ are elapsed time intervals since $ct = 0 = ct'$, according to $K$ and $K'$ respectively. While $c\Delta t'$ is a proper time interval in $K'$, $c\Delta t$ is a difference in readings from two clocks, one placed at the origin in $K$ and a second one at spatial position $x = V\Delta t$. Therefore $c\Delta t$ is simply a time coordinate or improper elapsed time as some people call it.

From here on we will be neglecting the y-direction in most of the situations.

Let us emphasize that the inertial frames $K$ and $K'$ have a common spacetime origin and that $K'$ has already the dimensionless speed $\beta(V)$ at $(x,ct) = (0,0) = (x',ct')$.

## IV - Doppler k-factor

Now let a light pulse be emitted at $(x,ct) = (0, c\Delta t_1)$ in $K$, along $x$-axis towards positive values of $x$. Suppose that at the spatial origin of $K'$ there is a mirror $M_3'$ that reflects back the light. The reflection occurs and at $(x,ct) = (V\Delta t_R, c\Delta t_R)$ under the viewpoint of observers in $K$ and at $(x',ct') = (0, c\Delta t_R')$ according to $K'$. The two time readings are related by

$$c\Delta t_R = \gamma(V) c\Delta t_R'. \tag{7}$$

Since all the clocks in $K$ are synchronized, the one which is side by side facing the clock in $K'$ is showing $c\Delta t_R$, the same exact value shown by the clock at the origin of $K$ while the clock in $K'$ shows $c\Delta t_R'$.

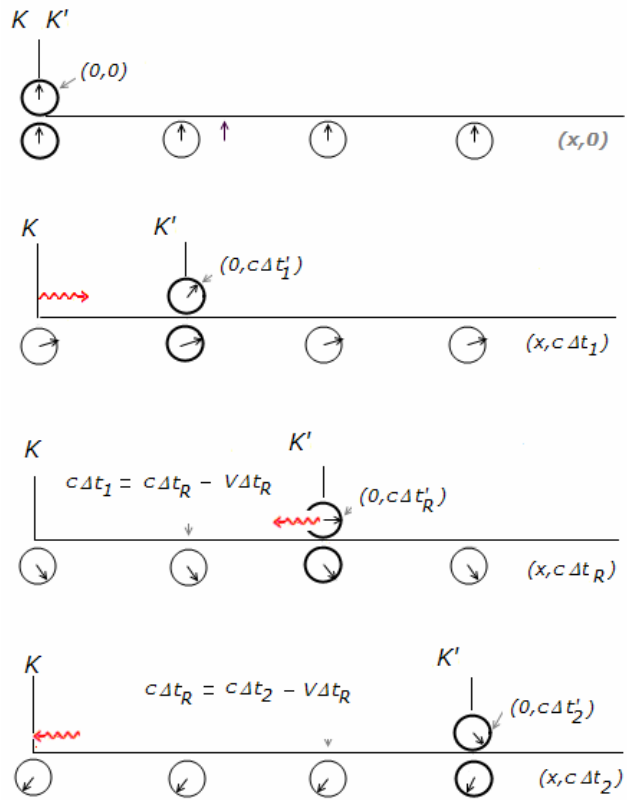

Figure 3: All clocks in $K$ are synchronized; $K'$ has only one clock, the light clock. A mirror $M_3'$, not shown, is attached to the spatial origin in $K'$ such that it can reflect to the left the light that comes from the spatial origin in $K$. Here we "see" the light ray emitted at $(x,ct) = (0, c\Delta t_1)$, its reflection at $(x,ct) = (V\Delta t_R, c\Delta t_R)$ and its return to $(x,ct) = (0, c\Delta t_2)$. At the moment of the emission, all the clocks shown the same time reading $c\Delta t_1$ whereas the clock in $K'$ shows $c\Delta t_1'$. The relation between these two values is $c\Delta t_1 = \gamma(V) c\Delta t_1'$. The time coordinates at the reflection are $c\Delta t_R = \gamma(V) c\Delta t_R'$ and at the return they are $c\Delta t_2 = \gamma(V) c\Delta t_2'$. Since light has constant speed no matter the direction it goes, it is easy to see that $c\Delta t_2 + c\Delta t_1 = 2c\Delta t_R$ and that $c\Delta t_2 - c\Delta t_1 = 2V\Delta t_R$.

After the reflection, the pulse reaches back the origin of $K$ at the event $(x,ct) = (0, c\Delta t_2)$. See Figure 3 for the corresponding scenario.

From the Figure 3 we can see promptly that

$$c\Delta t_1 = c\Delta t_R - V\Delta t_R \tag{8}$$

$$c\Delta t_2 = c\Delta t_R + V\Delta t_R. \tag{9}$$

From expressions (7), (8) and (9) we find

$$\frac{c\Delta t_2}{c\Delta t_R'} = \frac{c\Delta t_R + V\Delta t_R}{c\Delta t_R'} = (1 + \beta(V))\gamma(V) = k(V) \tag{10}$$

$$\frac{c\Delta t_R'}{c\Delta t_1} = \frac{c\Delta t_R'}{c\Delta t_R - V\Delta t_R} = \frac{1}{(1 - \beta(V))\gamma(V)} = k(V), \tag{11}$$

where $k(V)$ is Doppler $k$-factor [5,6]:

$$k(V) = \sqrt{\frac{1 + \beta(V)}{1 - \beta(V)}}. \tag{12}$$

## V - Addition of velocities

Now let us add a third light clock to our scenario. Let it be at rest at origin of inertial frame $K''$ and displayed exactly as the two previous ones; Mirrors $M_1''$ and $M_2''$ aligned along $y''$-axis. $K''$ is assumed to be moving with constant dimensionless speed $\beta(u)$ with respect to $K$, the movement being along the $x$-axis toward positive values of $x$. The initial conditions are the same as before: the three inertial frames have a common spacetime origin $(x,ct) = (0,0) = (x',ct') = (x'',ct'')$; $K'$ and $K''$ have already dimensionless speeds $\beta(V)$ and $\beta(u)$ with respect to $K$ when they meet at the origin.

In the previous section we used a thought experiment to obtain the Doppler k-factor. Let us do that experiment again, now with one more inertial frame ($K''$).

A light pulse is emitted in $K$, at $(x,ct) = (0,c\Delta t_1)$. At spatial origin of $K'$ there is a half-silvered mirror $M_3'$. Then part of the incident light is reflected back to the spatial origin of $K$ and part continues its path towards inertial frame $K''$. The event corresponding to the reflection by $M_3'$ is $(x,ct) = (V t_R, c\Delta t_R)$ as seen by observers in $K$, and $(x',ct') = (0,c\Delta t_R')$ by the observer in $K'$. The reflected pulse reaches the origin of $K$ at $(x,ct) = (0,c\Delta t_2)$. All the expressions from (7) to (12) follow naturally as before.

The part of the pulse which goes towards $K''$ is also reflected back by a mirror $M_3''$ placed at the spatial origin of $K''$. Then consider the following sequence: a light pulse is emitted at $(x,ct) = (0,c\Delta t_1)$, reflected at $(x,ct) = (u\Delta t_S, c\Delta t_S)$ by the mirror $M_3''$ in $K''$, and returned to the origin in $K$ at $(x,ct) = (0,c\Delta t_3)$ according to observers in $K$. To the observer in $K''$ this reflection occurs at $(x'',ct'') = (0, c\Delta t_S'')$. Again, since all the clocks in $K$ are synchronized, the one which is side by side facing the clock in $K''$ is showing $c\Delta t_S$, the same exact value shown by the clock at the origin of $K$ while the clock in $K''$ shows $c\Delta t_S''$. The two time readings are related by

$$c\Delta t_S = \gamma(u) c\Delta t_S'' \tag{13}$$

where

$$\gamma(u) = \frac{1}{\sqrt{1 - \beta(u)^2}} \tag{14}$$

By following the same steps from expression (8) to (12), we have

$$c\Delta t_1 = (1 + \beta(u))c\Delta t_S \tag{15}$$

$$c\Delta t_3 = (1 - \beta(u))c\Delta t_S \tag{16}$$

and from expressions (11), (13) and (16) we can write

$$\frac{c\Delta t_3}{c\Delta t_S{''}} = (1+\beta(u))\gamma(u) = k(u) \tag{17}$$

$$\frac{c\Delta t_S{''}}{c\Delta t_1} = \frac{1}{(1-\beta(u))\gamma(u)} = k(u). \tag{18}$$

where $k(u)$ is the Doppler k-factor between $K$ and $K''$:

$$k(u) = \sqrt{\frac{1+\beta(u)}{1-\beta(u)}} \tag{19}$$

Now there is the relative motion between $K'$ and $K''$ to analyze. Remember the half-silvered mirror that we assumed to exist at the spatial origin of $K'$. The pulse that comes from the origin of $K$ has part of it moving towards $K''$. Then we should consider the following sequence: a light pulse is "emitted" at $(x',ct') = (0,c\Delta t_R')$, reflected at $(x',ct') = (u'\Delta t_S',c\Delta t_S')$ and returned to the origin in $K'$ at $(x',ct') = (0,c\Delta t_4')$ according to observers in $K'$. Notice that we said *observers* in $K'$. At this point we have to admit the existence of an infinite number of observers in $K'$, each one with his (her) wristwatch. Then $c\Delta t_S'$ is the reading of the clock in $K'$ which is side by side with the one in $K''$. Since we assumed that all the clocks in $K'$ are synchronized according to Einstein's procedure, $c\Delta t_S'$ is the same reading as of the clock at the spatial origin of $K'$. Obviously the reflection occurs at $(x'',ct'') = (0,c\Delta t_S'')$ to the observer in $K''$.

Let $\beta(u')$ be the dimensionless speed of $K''$ with respect to $K'$. Then the relation between $c\Delta t_S''$ and $c\Delta t_S'$ is given by

$$c\Delta t_S' = \gamma(u') c\Delta t_S'' \tag{20}$$

$$\gamma(u') = \frac{1}{\sqrt{1-\beta(u')^2}}. \tag{21}$$

According to observers in $K'$ the measures $c\Delta t_S'$ and $c\Delta t_S''$ are simultaneous.

By following the same steps from expression (8) to (12), we get

$$c\Delta t_R' = (1+\beta(u'))c\Delta t_S' \tag{22}$$

$$c\Delta t_4' = (1-\beta(u'))c\Delta t_S' \tag{23}$$

$$\frac{c\Delta t_4'}{c\Delta t_S{''}} = (1+\beta(u'))\gamma(u') = k(u') \tag{24}$$

$$\frac{c\Delta t_S{''}}{c\Delta t_R'} = \frac{1}{(1-\beta(u'))\gamma(u')} = k(u'). \tag{25}$$

$$k(u') = \sqrt{\frac{1+\beta(u')}{1-\beta(u')}} \tag{26}$$

From expressions (11) and (25) we have

$$\frac{c\Delta t_S{''}}{c\Delta t_1} = \left(\frac{c\Delta t_S{''}}{c\Delta t_R'}\right)\left(\frac{c\Delta t_R'}{c\Delta t_1}\right) \tag{27}$$

which is the same as

$$k(u) = k(u')k(V) \tag{28}$$

or

$$\sqrt{\frac{1+\beta(u)}{1-\beta(u)}} = \sqrt{\frac{1+\beta(u')}{1-\beta(u')}}\sqrt{\frac{1+\beta(V)}{1-\beta(V)}}. \qquad (29)$$

Solving equation (29) for $\beta(u)$ we have the relativistic expression for the addition of velocities $\beta(u')$ and $\beta(V)$:

$$\beta(u) = \frac{\beta(u') + \beta(V)}{1 + \beta(u')\beta(V)}. \qquad (30)$$

## VI - Lorentz Transformations

Let us now focus our attention on the second reflection, the one which occurs at the mirror $M_3''$ placed at spatial origin of $K''$. The spacetime coordinates for such event are $(x'',ct'') = (0,c\Delta t_S'')$ in $K''$. In $K$ and $K'$ they are $(x,ct) = (\Delta x_S, c\Delta t_S)$ and $(x',ct') = (\Delta x_S', c\Delta_S t')$ respectively. According to what we have discussed before, the proper time $c\Delta t_S''$ appears to $K$ as time coordinate $c\Delta t_S$

$$c\Delta t_S = \gamma(u) c\Delta t_S'' \qquad (31)$$

and to $K'$ as time coordinate $c\Delta t_S'$

$$c\Delta t_S' = \gamma(u') c\Delta t_S''. \qquad (32)$$

We know that the spatial coordinate $\Delta x_S'' = 0$ since we are referring to the clock at the spatial origin in $K''$. What are the values for $\Delta x_S$ and $\Delta x_S'$? We will be looking for them.

Picking expressions (31) and (32) and eliminating $c\Delta t_S''$ we obtain

$$\frac{c\Delta t_S}{c\Delta t_S'} = \frac{\gamma(u)}{\gamma(u')}, \qquad (33)$$

where, by using expression (30) for the addition of speeds,

$$\frac{\gamma(u)}{\gamma(u')} = \gamma(V)(1 + \beta(u')\beta(V)). \qquad (34)$$

Then

$$c\Delta t_S = \gamma(V)(1 + \beta(u')\beta(V)) c\Delta t_S' \qquad (35)$$

Recall that $c\Delta t_S$ and $c\Delta t_S'$ *are not* proper time intervals; they are the elapsed time interval that observers in $K$ and $K'$ (at their viewpoint) attribute respectively to $c\Delta t_S''$ in $K''$. The quantities $c\Delta t_S$ and $c\Delta t_S'$ are simply time coordinate intervals; some people call them improper time intervals [7]. Expression (35) relates two non simultaneous elapsed non-proper time coordinates referring to the same events: $(x'',ct'') = (0,0)$ and $(x'', ct'') = (0,c\Delta t_S'')$.

The dimensionless speed of $K''$ with respect to $K'$ is $\beta(u')$, then

$$\Delta x_S' = \beta(u') c\Delta t_S'. \qquad (36)$$

So the expression (35) becomes

$$c\Delta t_S = \gamma(V)(c\Delta t_S' + \beta(V)\Delta x_S') \qquad (37)$$

The dimensionless speed of $K''$ with respect to $K$ is $\beta(u)$, then

$$\Delta x_S = \beta(u) c\Delta t_S. \qquad (38)$$

Multiplying both sides of expression (35) by $\beta(u)$ we get

$$\beta(u)\, c\Delta t_S = \left(\frac{\gamma(u)}{\gamma(u')}\beta(u)\right) c\Delta t_S'. \tag{39}$$

By using (30) and (34) we have

$$\beta(u)\, c\Delta t_S = \left(\gamma(V)(1+\beta(u')\beta(V))\right) c\Delta t_S' \tag{40}$$

$$\Delta x_S = \gamma(V)(\beta(u')+\beta(V))\, c\Delta t_S' \tag{41}$$

$$\Delta x_S = \gamma(V)\left(1+\frac{\beta(V)}{\beta(u')}\right)\Delta x_S' \tag{42}$$

This expression relates the spatial distance $\Delta x_S$ traveled by the third light clock since $(x,ct) = (0,0)$ up to $(x,ct) = (u\Delta t_S, c\Delta t_S)$, according to observers in $K$; $\Delta x_S'$ is the spatial distance traveled by the third light clock since $(x',ct') = (0,0)$ up to $(x',ct') = (u'\Delta t_S', c\Delta t_S')$ at the viewpoint of $K'$.

Expression (33) can also be written as

$$\Delta x_S = \gamma(V)(\Delta x_S' + \beta(V)\, c\Delta t_S') \tag{43}$$

Notice that expressions (37) and (43) are just the Lorentz Transformations connecting spatial $\Delta x_S'$ and time coordinate $c\Delta t_S'$ intervals in $K'$ to spatial $\Delta x_S$ and time $c\Delta t_S$ intervals in $K$. Since $(\Delta x_S, c\Delta t_S)$ is arbitrary in the same way $(0, c\Delta t_1)$ was arbitrary, we can drop the subscript S and simply write

$$\Delta x = \gamma(V)(\Delta x' + \beta(V)\, c\Delta t') \tag{44}$$

$$c\Delta t = \gamma(V)(c\Delta t' + \beta(V)\Delta x') \tag{45}$$

*These equations are completely symmetrical between the space and time coordinates.* Interchanging everywhere $\Delta x$ and $c\Delta t$ leaves the equations invariant. This is a manifestation of the striking symmetry that exists between space and time in special relativity; $\Delta x$ and $c\Delta t$ are intimate connected. This is why it is so appropriate to speak of a single entity named spacetime. In Galilean relativity, space and time are totally distinct.

## VII - Minkowski diagram

When we use spacetime Minkowski diagram representation we see that it is easier to visualize the properties we have been discussing. See Figure 4 for the scenario of our discussion with the help of the Minkowski diagram. There we have the three inertial frames mentioned before: $K$ at rest (assumed stationary), $K'$ moving with constant dimensionless speed $\beta(V)$ along the *x-axis* and $K''$ moving with constant dimensionless speed $\beta(u)$ with respect to $K$ also along *x-axis*. Each of the frames has a light clock in which light pulses back and forth along the *y-axis* directions; the *y-axis* direction is not shown. As we said all the three inertial frames have a common spacetime origin: $(x,ct) = (x',ct') = (x'',ct'') = (0,0)$.

In the Figure 4 we "see" the light ray emitted at $(x,ct) = (0, c\Delta t_1)$, its reflection at $(x,ct) = (V\Delta t_R, c\Delta t_R)$ and its return to $(x,ct) = (0, c\Delta t_2)$. The half-silvered mirror in the origin of $K'$ allows part of the light ray to propagate from $(x,ct) = (V\Delta t_R, c\Delta t_R)$ until $(x,ct) = (u\Delta t_S, c\Delta t_S)$ and then after reflection it reaches first $(x,ct) = (V\Delta t_4, c\Delta t_4)$ and finally $(x,ct) = (0, c\Delta t_3)$. All horizontal lines are simultaneity lines under the viewpoint of observers in $K$. The horizontal line segment connecting the event $(x,ct) = (0, c\Delta t_S)$ to the event $(x,ct) = (V\Delta t_S, c\Delta t_S)$ is the spatial distance between them: $\Delta x_S = V\Delta t_S$. The relation between the time coordinate interval $\Delta t_S$ and the proper time interval $\Delta t_S''$ is

$$c\Delta t_S = \gamma(u)\, c\Delta t_S'' \tag{46}$$

The line connecting the event $(x',ct') = (0,c\Delta t_S')$ to event $(x',ct') = (u'\Delta t_S',c\Delta t_S')$ – simultaneous events to $K'$ - is parallel to the $O'X'$ axis and it is the spatial distance between the mentioned events: $\Delta x_S' = u'\Delta t_S'$. The relation between the time coordinate interval $\Delta t_S'$ and the proper time interval $\Delta t_S''$ is

$$c\Delta t_S' = \gamma(u')c\Delta t_S'' \tag{47}$$

Now, by following the same steps between expression (31) and (37) we get

$$c\Delta t_S = \gamma(V)(c\Delta t_S' + \beta(V)\Delta x_S') . \tag{48}$$

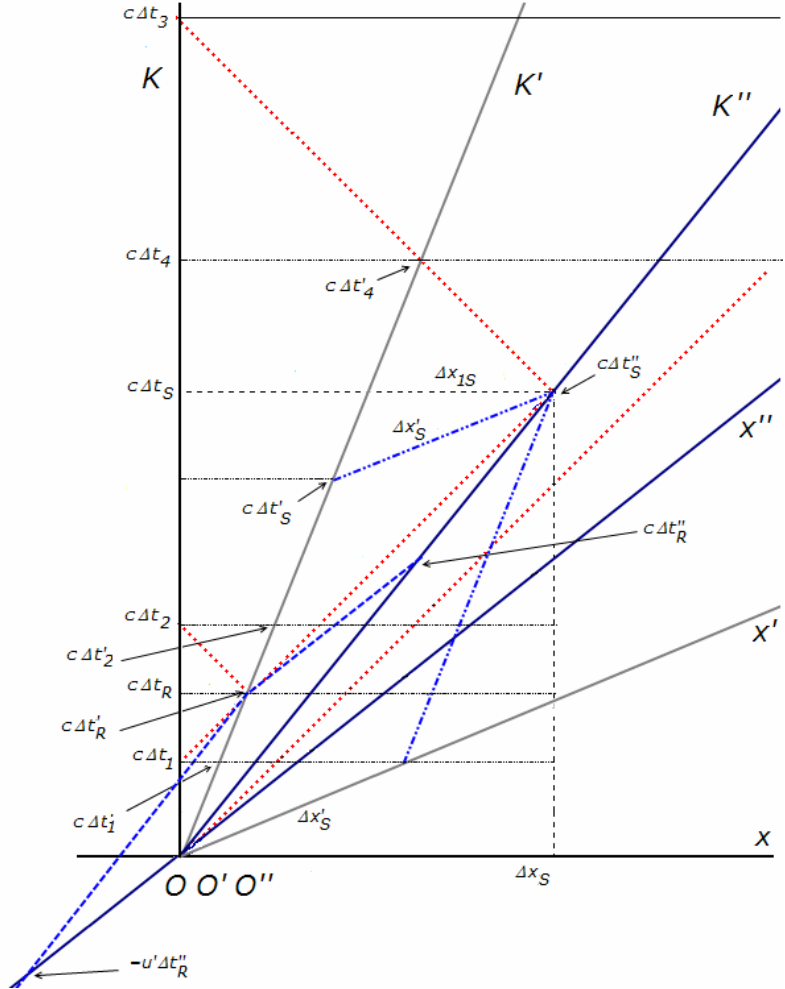

**Figure 4:** Here we "see" the light ray emitted at $(x,ct) = (0, c\Delta t_1)$, its reflection at $(x,ct) = (V\Delta t_R, c\Delta t_R)$ and its return to $(x,ct) = (0, c\Delta t_2)$. The half-silvered mirror in $K'$ allows part of the light ray to propagate to $(x,ct) = (u\Delta t_S, c\Delta t_S)$ and then after reflection it reaches first $(x,ct) = (V\Delta t_4, c\Delta t_4)$ and finally $(x,ct) = (0, c\Delta t_3)$. Horizontal lines are simultaneity lines for observers in $K$, while parallel line to $x'$-axis is a simultaneity line to $K'$ and the parallel line to $x''$-axis is a simultaneity line to $K''$.

Observe now that the horizontal segment line $\Delta x_{1S}$ can be expressed as

$$\Delta x_{1S} = (\beta(u) - \beta(V))c\Delta t_S = (\beta(u) - \beta(V))\frac{\gamma(u)}{\gamma(u')}c\Delta t_S' \tag{49}$$

$$\Delta x_{1S} = \left(\frac{\beta(u) - \beta(V)}{\beta(u')}\right) \frac{\gamma(u)}{\gamma(u')} \Delta x_S' \tag{50}$$

By using expression (30) for addition of speeds we have

$$\Delta x_{1S} = \frac{1}{\gamma(V)} \Delta x_S', \tag{51}$$

which is just the relativistic length contraction formula. The proper spatial distance $\Delta x_S'$ in $K'$ is "seen" in $K$ as $\Delta x_{1S} < \Delta x_S'$.

With the result (51) and expression (34), the spatial coordinate interval $\Delta x_S$ can be put as

$$\Delta x_S = \beta(V) c\Delta t_S + \frac{\Delta x_S'}{\gamma(V)} = \frac{\gamma(u)}{\gamma(u')} c\Delta t_S' + \frac{\Delta x_S'}{\gamma(V)}, \tag{52}$$

$$\Delta x_S = \gamma(V)(1 + \beta(u')\beta(V)) c\Delta t_S' + \frac{\Delta x_S'}{\gamma(V)} \tag{53}$$

$$\Delta x_S = \gamma(V)(1 + \beta(V)\beta(u')) \beta(V) c\Delta t_S' + \frac{\Delta x_S'}{\gamma(V)} \tag{54}$$

$$\Delta x_S = \gamma(V)\beta(V)\left(c\Delta t_S' + \beta(V)\beta(u') c\Delta t_S' + \frac{\Delta x_S'}{\beta(V)\gamma(V)^2}\right) \tag{55}$$

Since

$$\Delta x_S' = \beta(u') c\Delta t_S' \tag{56}$$

the expression for $\Delta x_S$ can finally written as

$$\Delta x_S = \gamma(V)\left(\left(\beta(V)^2 + \frac{1}{\gamma(V)^2}\right)\Delta x_S' + \beta(V) c\Delta t_S'\right) \tag{57}$$

or

$$\Delta x_S = \gamma(V)(\Delta x_S' + \beta(V) c\Delta t_S'). \tag{58}$$

This equation and equation (48) (dropping the subscript $S$) are the standard Lorentz Transformations

$$\Delta x = \gamma(V)(\Delta x' + \beta(V) c\Delta t') \tag{59}$$

$$c\Delta t = \gamma(V)(c\Delta t' + \beta(V)\Delta x'). \tag{60}$$

## VIII - Spacetime Interval Invariance

If we take the Lorentz Transformations (59) and (60) and calculate $(c\Delta t)^2 - (\Delta x)^2$ we find a quantity $\Delta s^2$ that does not depend on which frame the observer is. The quantity $\Delta s$ is the spacetime interval between the events referred by $c\Delta t$ and $\Delta x$, plays the role of a spacetime distance and it is invariant:

$$\Delta s^2 = c^2 \Delta t^2 - \Delta x^2 = c^2 \Delta t'^2 - \Delta x'^2 \tag{61}$$

As an example, consider, for instance, the first reflection event in our scenario. Its coordinates are $(x,ct) = (V\Delta t_R, c\Delta t_R)$ in $K$, $(x',ct') = (0, c\Delta t_R')$ in $K'$ and $(x'',ct'') = (-u'\Delta t_R'', c\Delta t_R'')$ in $K''$. Easily, we get that

$$c^2 \Delta t_R^2 - \Delta x_R^2 = (1 - \beta(V)^2) c^2 \Delta t_R^2 = \frac{1}{\gamma(V)^2} c^2 \Delta t_R^2 \tag{62}$$

$$c^2 \Delta t_R'^2 - \Delta x_R'^2 = c^2 \Delta t_R'^2 \tag{63}$$

$$c^2 \Delta t_R''^2 - \Delta x_R''^2 = (1 - \beta(u')^2) c^2 \Delta t_R''^2 = \frac{1}{\gamma(u')^2} c^2 \Delta t_R''^2 \tag{64}$$

Since

$c\Delta t_R = \gamma(V) c\Delta t_R'$ and $c\Delta t_R'' = \gamma(u') c\Delta t_R'$, the three expressions above are equal to the square of the proper time $c\Delta t_R'$, the value that the observer at rest in $K'$ reads in his(her) wristwatch. Then the spacetime interval for first reflection event is $s = s_R = c\Delta t_R'$.

One more example: pick the second reflection event. Its coordinates are $(x,ct) = (u\Delta t_S, c\Delta t_S)$ in $K$, $(x',ct') = (u'c\Delta t_S', c\Delta t_S')$ in $K'$ and $(x'',ct'') = (0, c\Delta t_R'')$ in $K''$. Now, we get that

$$c^2 \Delta t_S^2 - \Delta x_S^2 = (1 - \beta(u)^2) c^2 \Delta t_S^2 = \frac{1}{\gamma(u)^2} c^2 \Delta t_S^2 \tag{65}$$

$$c^2 \Delta t_S'^2 - \Delta x_S'^2 = (1 - \beta(u')^2) c^2 \Delta t_S'^2 = \frac{1}{\gamma(u')^2} c^2 \Delta t_S'^2 \tag{66}$$

$$c^2 \Delta t_S''^2 - \Delta x_S''^2 = c^2 \Delta t_S''^2. \tag{67}$$

We can straightforwardly obtain the spacetime interval for the second reflection event which is the invariant quantity $s = s_S = c\Delta t_S''$.

## IX - Conclusions

We have started with the light clock to derive time dilation, Doppler k-factor, addition of velocities, length contraction, Lorentz Transformations, invariance of spacetime interval. We specially tried to show that it is not difficult to deal with the spacetime coordinates in special relativity by manipulating their interrelations and interpreting a few simple events with the help of Minkowski diagram.

[1] H. Minkowski, *"Space and Time"*, Cologne Conference, September 21, 1908 (reprint in English, "The Principle of Relativity", (Dover Publications, New York, 1923), p.75-91.
[2] A. Einstein, "Zur Elektrodynamik der bewegter Körper", Ann. Phys., **17**, 891 (1905) (reprint in English, *"The Principle of Relativity"*, Dover Publications, New York, 35-65).
[3] N. D. Mermin, *"Space and Time in Special Relativity"* (McGraw-Hill, New York, 1968)
[4] B. L. Coulter, "Relativistic motion in a plane", Am. J. Phys. **48**, 633 (1980).
[5] H. Bondi, *"Relativity and Common Sense"* (Dover Publications, New York, 1962)
[6] N. D. Mermin, "An introduction to spacetime diagrams", Am. J. Phys. **65**, 477-486 (1997).
[7] L. Sartori, *Understanding Relativity - A Simplified Approach to Einstein's Theories* (University of California Press, London, England, 1996)